\documentclass[
    ,final            
  ]
  {aipproc}

\layoutstyle{6x9}

\begin{document}

\title{The Moderate Cooling Flow Model and Feedback in Galaxy Formation}


\author{Noam Soker}{
  address={Dept. of Physics, Technion, Haifa 32000 Israel; soker@physics.technion.ac.il}
}

\author{Assaf Sternberg}{
  address={Dept. of Physics, Technion, Haifa 32000 Israel}
}
\author{Fabio Pizzolato}{
  address={Osservatorio Astronomico di Brera, Via Brera 28,  20121 Milano}
}

\begin{abstract}
For the recent four years we have been studying feedback heating in
cooling flow (CF) clusters by AGN activity that inflate bubbles by jets;
this short contribution to a meeting summarizes our main results.
To achieve our results we had to self-consistently inflate
the bubbles with jets, rather than inject them artificially.
Our main results are as follows
 (1) Feedback mechanisms that are based on Bondi accretion fail.
Instead, the accretion to the central super-massive black hole (SMBH) is
in the form of cold dense blobs that fall-in from an extended region.
(2) Slow massive wide (SMW) jets, or rapidly precessing jets, can
inflate bubbles similar to those observed in CF clusters.
(3) Contrary to some claims in the literature, the inflated bubbles are stable
   for a relatively long time, becoming unstable only at later times.
(4) A single bubble inflation episode excites multiple sound waves and shocks.
 These can then heat the intracluster medium (ICM).
(5) Mixing of the bubble material to the ICM is efficient, and can serve as a main
heating channel.
(6) The heating processes work in all directions, and can explain the heating
of the ICM in CF in clusters and in galaxies.
\end{abstract}

\maketitle


\section{Accretion to the super-massive BH}

\textbf{(1) The failure of the Bondi accretion feedback}
\newline
 In several papers and talks we showed that the Bondi accretion fails
to explain feedback heating in cooling flows (CFs) in clusters of galaxies (Soker 2006).
For more arguments the reader should consult that paper.
Here we only point to the two main reasons for the failure of a feedback heating based on the
Bondi accretion.

\emph{(i) There is no time for the feedback to work.}
In the Bondi accretion process the inflow velocity at the Bondi radius
\begin{equation}
R_B=\frac {2GM_{BH}}{C_s^2}= 35
\left( \frac{M_{BH}} {10^9 M_\odot } \right)
\left( \frac{C_s} {500 ~{\rm km}~{\rm s}^{-1} } \right)^{-2}
{\rm pc},
\label{eq:rb1}
\end{equation}
is about the sound speed $C_s \sim 500 ~{\rm km}~{\rm s^{-1}}$, with typical values for CF clusters.
From mass conservation the inflow velocity varies as $r^{-2}$ (because the ICM density profile is
flat in the inner regions of clusters), and its typical
value at $r \sim 1 ~{\rm kpc}$ is $v_{\rm in} \sim 1 ~{\rm km}~{\rm s}^{-1}$.
The time it takes for the gas to flow inward from a distance of $r=1 {\rm kpc}$
and feed the SMBH is
\begin{equation}
\tau_{FB-BH}(1 ~{\rm kpc}) \simeq  \frac{1 {\rm kpc}}{1 ~{\rm km}~{\rm s}^{-1}} =
10^9 {\rm yr}.
\label{eq:tau1}
\end{equation}
This time is longer than the cooling time of most CF clusters at
$r \sim 1 ~{\rm kpc}$. The situation at $r > 3~ {\rm kpc}$ is hopeless, as the inflow time
is more than the age of the cluster.
\newline
\emph{(ii) Insufficient heating in clusters. }
Most supporters of feedback by Bondi accretion cite the paper by Allen et al. (2006).
However, Allen et al. (2006), noted that their results cannot be extrapolated
to CFs in cluster of galaxies.
Most likely, the Bondi accretion process cannot explain the feedback heating in the
process of galaxy formation as well.
\newline
\textbf{(2) The cold feedback mechanism}
\newline
In the cold-feedback model (Pizzolato \& Soker 2005; Soker 2006, 2008b)
mass accreted by the central SMBH originates in non-linear over-dense blobs of gas
residing in an extended region of $r \sim 5-50 ~{\rm kpc}$; these blobs are
originally hot, but then cool faster than their environment and sink toward the center
(see also Revaz et al. 2008).
The mass accretion rate by the central black hole is determined by the cooling time
of the ICM, the entropy profile, and the presence of inhomogeneities (Soker 2006).
Most important, the ICM entropy profile must be shallow for the
blobs to reach the center as cold blobs.
Wilman et al. (2008) suggest that the behavior and properties of the cold clumps they
observe in the cluster A1664 support the cold feedback mechanism.

The cold feedback mechanism has the following consequences.
\newline
{(1)} Cooling flows do exist, but at moderate mass cooling rates.
            The \emph{moderate cooling flow model}  (Soker et al. 2001).
\newline
{(2)} The cold feedback mechanism explains why real clusters depart from
   an `ideal' feedback loop that is 100\% efficient in suppressing cooling and star formation.
   Simply, the feedback requires that non-negligible quantities of mass cool to low temperatures.
   Part of the mass falls to small radii.
\newline
{(3)} Part (likely most) of the inflowing cold gas is ejected back from the very inner region.
   This is done by the original jets blown by the SMBH (Soker 2008a). The ejection of this
   gas is done in a slow massive wide (SMW) bipolar outflow, which are actually two jets.
   The most striking example of an SMW outflow is presented in the seminal work of Moe et al.
(2009 ApJ submitted).
By conducting a thorough analysis, Moe et al. (2009) find the outflow
from the quasar SDSS J0838+2955 to have a velocity of $\sim 5000 ~{\rm km}~{\rm s}^{-1}$,
and a mass outflow rate of $\sim 600 M_\odot ~{\rm yr}^{-1}$,
assuming a cover fraction of $\delta \simeq 0.2$.
The cooling and ejection back of large quantities of mass, make the feedback process
not only of energy (heating), but also of mass (Pizzolato \& Soker 2005).
\newline
{(4)} As we showed in a series of five papers, such SMW jets can inflate the `fat' bubbles
that are observed in many CFs, in clusters, groups of galaxies, and elliptical galaxies.
The implication is that narrow radio jets that are observed in many cases are not
the main source of the energy inflating the bubbles.
We elaborate on this in the last section.
\newline
{(5)} The same mechanism that form SMW jets in CF clusters and galaxies,
can expel large quantities of mass during galaxy formation, and might
explain the SMBH-bulge mass correlation (Soker 2009).

\section{Inflating bubbles in cooling flows}

In a series of five papers we conducted 2.5D hydrodynamical numerical simulations,
and for the first time managed to systematically inflate `fat' bubbles
(i.e., large and almost spherical bubbles) as observed in CF clusters.
Main results of these papers are as follows.
\newline
\textbf{(1)}
Fat bubbles can be inflated by SMW jets (Sternberg et al. 2007; Sternberg \& Soker 2008b, 2009a,b).
Typical values are:  initial jet speed$\simeq 0.01-0.1c$;
mass loss rate into the two jets$\sim 1-100 M_\odot ~{\rm yr}^{-1}$;
a large half opening angle of $\alpha > 30^\circ$.
\newline
\textbf{(2)} Rapidly precession narrow jets have the same effect as wide jets
(Sternberg \& Soker 2008a, 2009a), and can also explain the formation of `fat' bubbles.
\newline
\textbf{(3)} Bubbles are known to be stable during the inflation phase
(Pizzolato \& Soker 2006). The reason is that the bubble's boundary is decelerated by the denser
ICM gas, and hence it is actually Rayleigh-Taylor  \emph{stable} during the inflation phase.
When we inflate bubbles self-consistently with jets
(as it should be done, and as we did in all our simulations), we further find that
the bubbles stay intact for a much longer time after the inflation phase ends
(Sternberg \& Soker 2008b).  
Basically, the dense shell moving outward ahead of the bubble (formed during
the inflation phase) and the vortexes formed inside the bubbles, stabilize the bubble.
These effects cannot be studied with bubbles that are artificially injected to the ICM.
\newline
\textbf{(4)} Every bubble inflation episode excites multiple sound waves and shocks
(Sternberg \& Soker 2009a). Such sound waves are observed in
some CF clusters, most clearly in Perseus (Fabian et al. 2006).
\newline
\textbf{(5)} Mixing of very-hot bubble material with hot ICM gas can be a major channel to
heat the ICM (Sternberg \& Soker 2008b, 2009a; also Br\"uggen et al. 2009).
\newline
\textbf{(6)} The last two processes are effective not only in the directions of propagation
of the jets and bubbles, but also perpendicular to this symmetry axis.
It seems that these processes provide the energy channel from the jets to the ICM
in all relevant regions. Again, self-consistent inflation of bubbles by jets is required
to show that.
\newline
\textbf{(7)} Narrow very fast (even relativistic) jets can have negligible
energy content and still fill the bubbles with radio radiation.
We elaborate on this preliminary result next.

\section{IMPERFECT CORRELATION OF RADIO AND X-RAY BUBBLES}

In many clusters there is an imperfect morphological correlation between radio emission and deficient
X-ray bubbles. Examples were given in the talks by  Fabio Gastaldello and Simona Giacintucci.
More examples are:
(1) In NGC~5044 the radio emission fills only one of three bubbles (David et al. 2009);
(2) Hercules A does not show X-ray holes coincident with the radio lobes
(Gizani \& Leahy 2004; Nulsen et al. 2005);
(3) In Abell 2626 no X-ray deficient bubbles are observed
at the locations of strong radio emission (Rizza et al. 2000; Wong et al. 2008).
We are also motivated  by that the energy required to inflate
bubbles is larger than the energy in radio emission (e.g., De Young 2006).

We note that weak radio emission at long wavelength can fill the bubbles. This is expected
as old radio jets might have filled the region.
Here we are interested in the radio morphology of recently launched jets.

The simulations were performed using the \emph{Virginia Hydrodynamics-I}
code (VH-1; Blondin et al. 1990), as described in our previous papers.
For the first 100~Myr time period of our simulation we used the parameters of Model I
(as described in Sternberg \& Soker 2009b), at the end of which a fat bubble has been
inflated.
 All detail are in that paper.

We study a three-dimensional axisymmetric flow with a 2D grid (referred to
as 2.5D).
The fast `radio' jet was turned on at the same time the SMW jet was turned off.
While the SMW jet was ejected along the symmetry axis (horizontal axis in the figures),
the fast jet was injected at an angle of $30^\circ$ to the symmetry axis.
In our 2.5D numerical code, this physically corresponds to a jet that is rapidly rotating
around the symmetry axis at an angle of $30^\circ$ (see Sternberg \& Soker 2008a).
The fast jet has a half opening angle of $10^\circ$, an injection velocity of
$v_f=75C_s=8.5 \times 10^4 ~{\rm km}~{\rm s^{-1}}$, a mass loss rate per two jet of
$\dot{M}_{2f}=10 M_\odot {\rm yr}^{-1}$,
and a total power of the two fast jets of $\dot E_{2f}=2.3 \times 10^{46} ~{\rm erg}~{\rm s}^{-1}$,
which is equal to the power of the SMW jets that inflated the bubbles up to $t=100~$Myr.

In reality, we would like the fast (radio) jets power to decline from about the power of the SMW jets
to $< 0.1$ times the power of the SMW jets, and be narrow jets with slow precession.
Such a jet covers a solid angle of $2 \pi (1-\cos 10^\circ)=0.03 \pi$.
 However, what matters for the penetration of the jets through the bubble is
the momentum flux (equals to the ram pressure).
In our 2.5D code one jet we inject covers a solid angle of $2 \pi (\cos 20^\circ-\cos40^\circ)=0.347 \pi$.
To maintain the desired momentum flux we had to use a power larger by a factor of
$\sim 0.347/0.03 \simeq 10$.
The evolution of the fast (radio) jet, as shown in the figures, teaches us
about the propagation of the fast jets and the morphology of
their shocked gas, but the total volume filled by the shocked fast jet's material is
larger than what a more realistic full 3D code will give.

 Our code does not explicitly separate the material originating in the SMW and the fast jets.
We identify the material originating in the fast jet by its high postshock temperature,
$ T > 5\times 10^{9}~$K.
In reality, this is where we expect to detect strong radio emission.
\begin{figure}
\begin{tabular}{ccc}
\hskip 0.2 cm
{\includegraphics[height=.19\textheight]{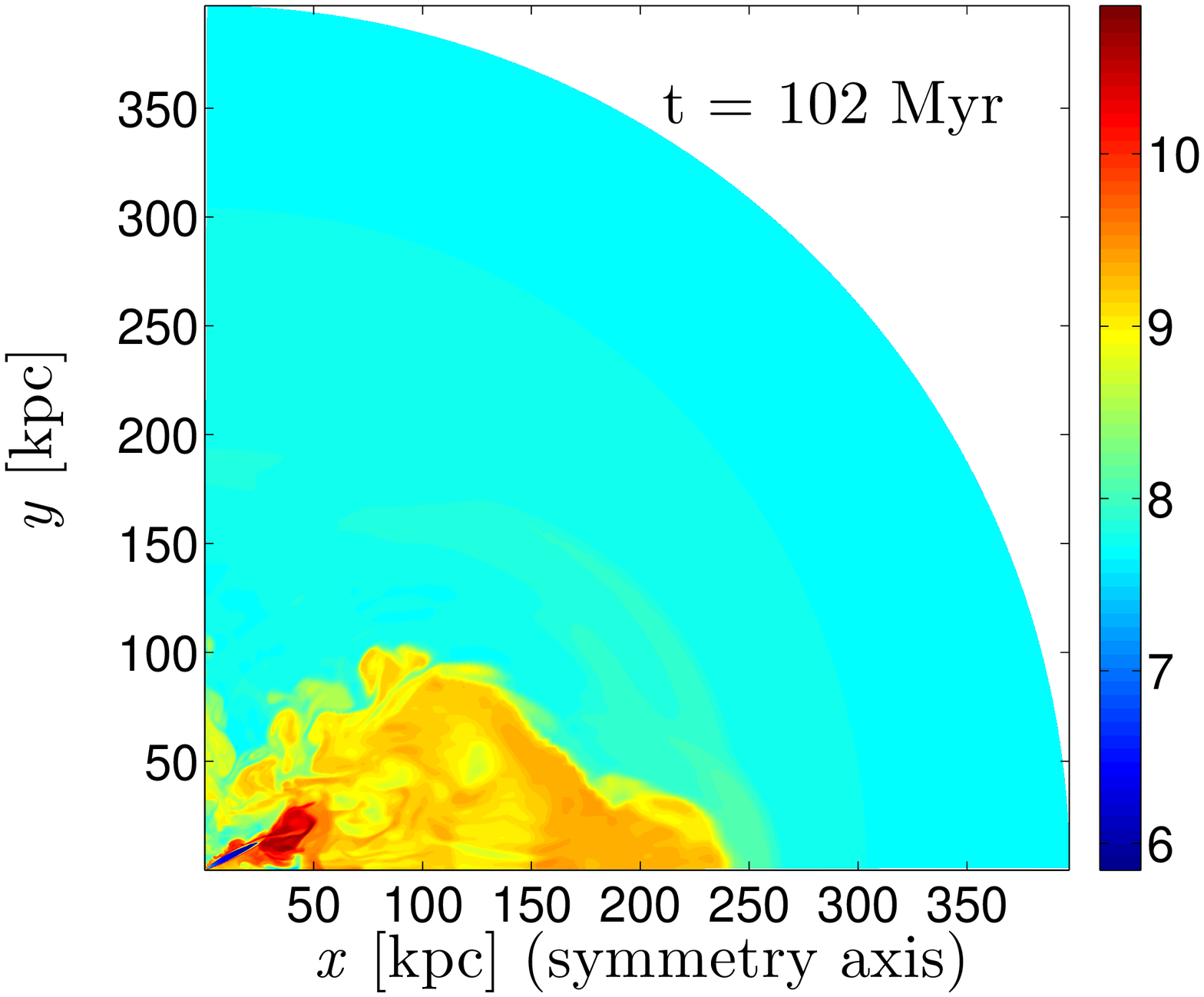}} &
\hskip -.70 cm
{\includegraphics[height=.19\textheight]{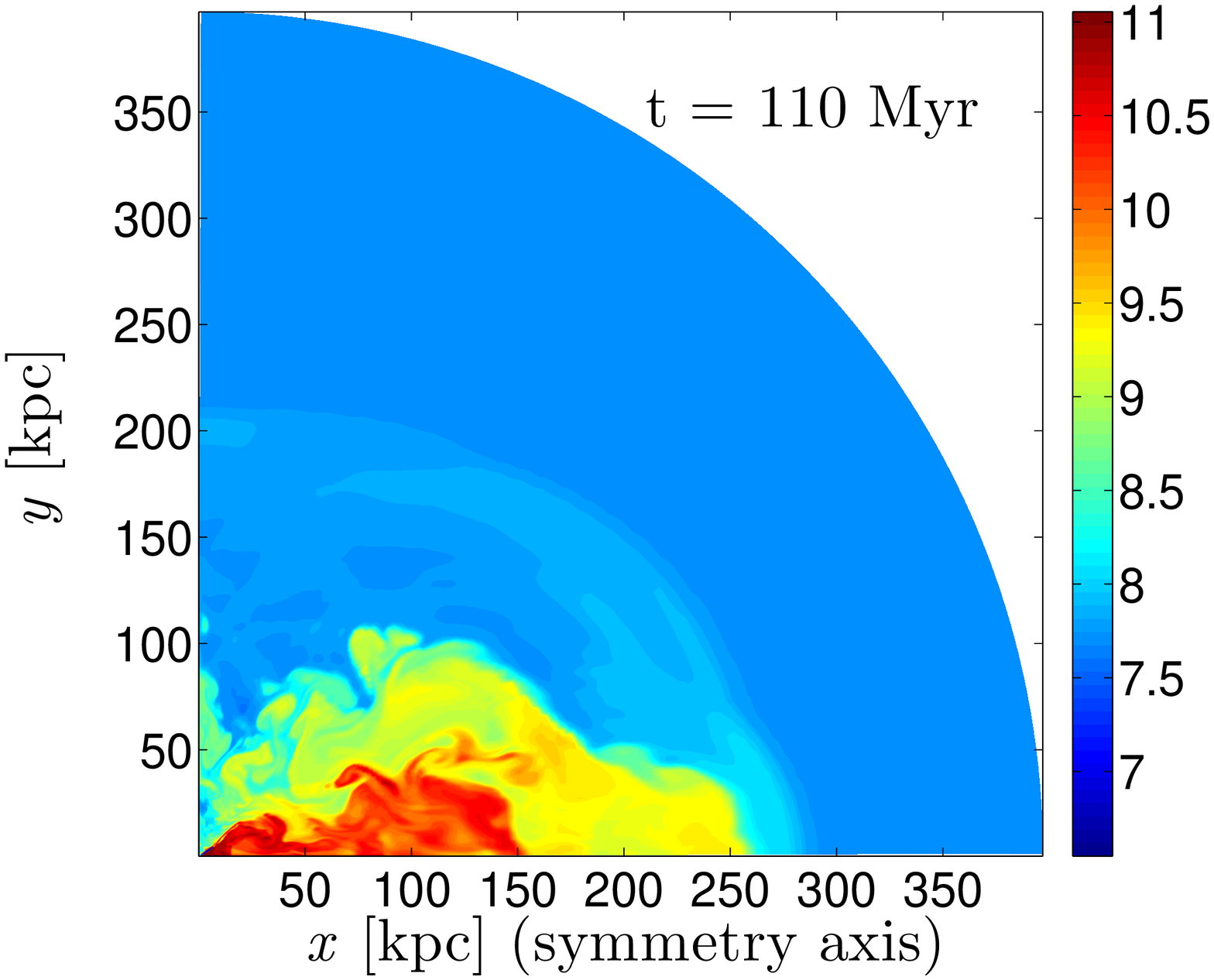}} &
\hskip -.70 cm
{\includegraphics[height=.19\textheight]{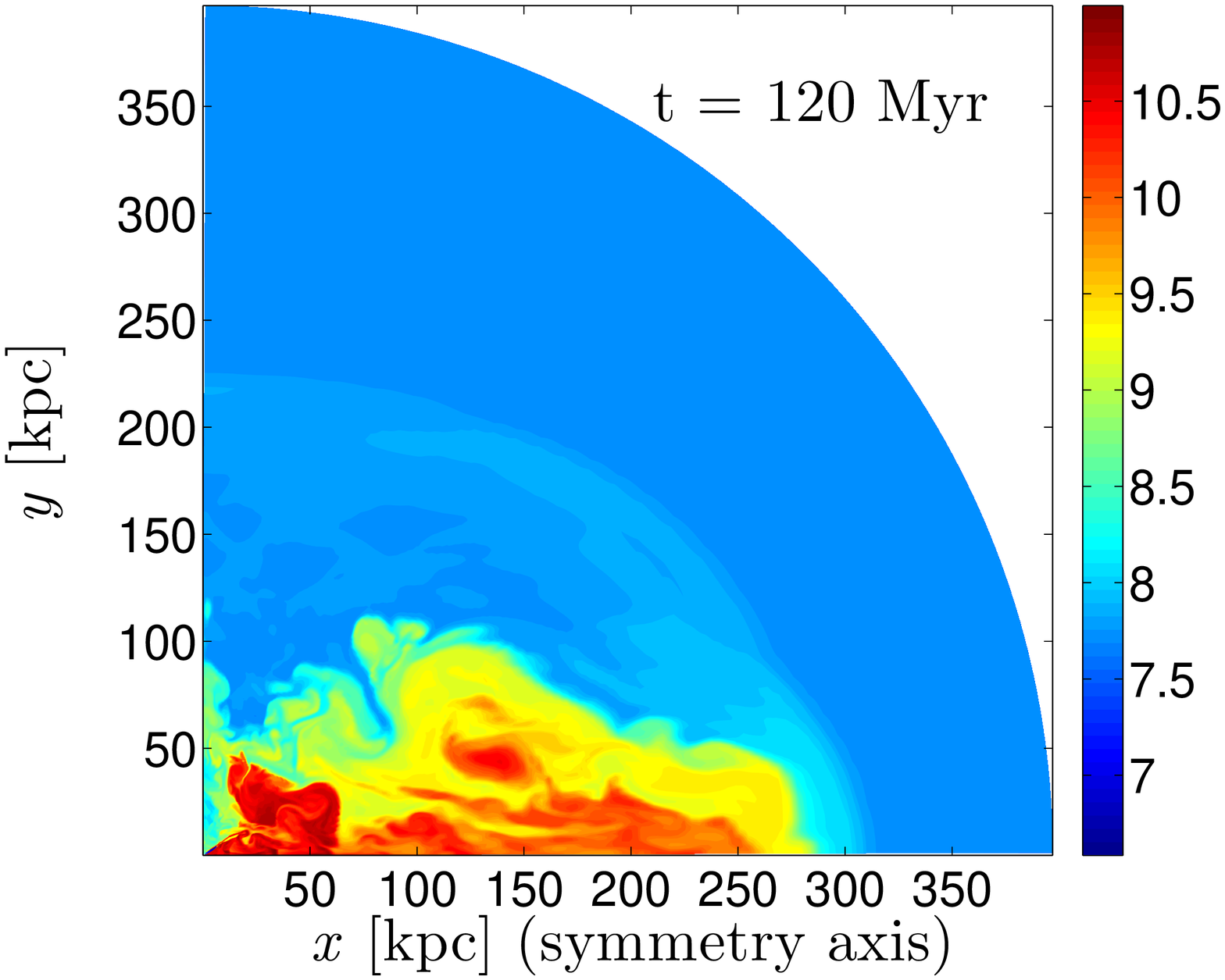}} \\
\end{tabular}
\caption{Log-temperature maps of the evolution of the fast `radio' jet.
The symmetry axis is along the $x$ (horizontal) axis. The $y$ axis is the
equatorial plane.
For $100~$Myr the slow massive wide (SMW) jet inflated the bubble as in Model~1
of Soker \& Sternberg (2009b).
The fast jet was turned on at $t=100~$Myr. Its postshock material is identified by the
regions of high-temperature of $T > 5\times 10^{9}~$K. These regions are where
strong radio emission is expected.
The color scaling is given in unites of $\log T(K)$.}
\label{fig:fast11}
\end{figure}

We note that the shocked fast jet material is spreading inside a large volume of the bubble.
However, it is not smooth, but has shapes of loops and arcs, as observed in some cases.
Namely, there is no one-to-one (perfect) correlation between the shocked fast
(`radio') jet material and the bubble.
The main reason is that it is injected in one direction, and that
the internal velocities of the gas inside the already inflated bubbles spreads the
shocked fast wind gas around.
In other words, the internal motion inside the already existing bubble spreads the shocked fast
jet material to large distances and to different directions.
This might give the (wrong) impression that the bubble was inflated by the fast jet.

The justification for the transition from SMW jets to much faster narrow jets can
be explained in the following way in the frame of the cold feedback model.
After most of the cold gas was expelled from the vicinity of the SMBH in one
of the feedback cycles, two changes occur.
First, the accretion rate declines and so is the power of the jets.
Second, the fast narrow jets launched by the SMBH do not encounter large
quantities of dense gas, that otherwise might prevent them from propagating
to large distances (Soker 2008a).
Therefore, instead of SMW jets we observe the narrow radio jets.
In reality the transition will be gradual, and both types of jets might coexist.
We propose this as an explanation to many cases where no perfect correlation
exist between the morphologies of strong radio emission and X-ray deficient bubbles.
If true, this implies that the observed radio jets are not the main energy source of the bubbles.








\bibliographystyle{aipprocl} 


\end{document}